\documentclass{ragtime}

\title[Energy dissipation in astrophysical simulations]%
   {Energy dissipation in astrophysical simulations: results of the Orszag-Tang test problem}
\author[F. Kayanikhoo, et al.
]%
{Fatemeh Kayanikhoo\at[]{1,a},
Miljenko \v{C}emelji\'{c} \at[]{1,2,3},\splitauthors
Maciek Wielgus \at[]{4},
Włodek Klu{\'z}niak \at[]{1}\\
\ins{1}Nicolaus Copernicus Astronomical Center,\splitins[1]
 Bartycka 18, 00-716, Warszaw, Poland,
 \\
\ins{2}Research Centre for Computational Physics and Data Processing,\\
Institute of Physics, Silesian University
in Opava,\splitins[1] Bezru\v{c}ovo n\'am.~13, CZ-746\,01 Opava,
Czech Republic,\\
\ins{3}Academia Sinica, Institute of Astronomy and Astrophysics,\splitins[1] P.O. Box 23-141,
Taipei 106, Taiwan\\
\ins{4}Max-Planck-Institut f\"ur Radioastronomie,\splitins[1] Auf dem H\"ugel 69, D-53121 Bonn, Germany\\
\ins{a}\Email{fatima@camk.edu.pl}}
\coentry{F, Kayanikhoo, et al.}%
{Energy dissipation in astrophysical simulations}

\providecommand{\pluto }{\texttt{PLUTO}}
\providecommand{\koral }{\texttt{KORAL}}

\newcommand{\beq}{\begin{equation}}
\newcommand{\beqa}{\begin{eqnarray}}
\newcommand{\eeq}{\end{equation}}
\newcommand{\eeqa}{\end{eqnarray}}
\begin{document}

\begin{abstract}
The magnetic field through the magnetic reconnection process affects the dynamics and structure of astrophysical systems. Numerical simulations are the tools to study the evolution of these systems. However, the resolution, dimensions, resistivity, and turbulence of the system are some important parameters to take into account in the simulations. In this paper, we investigate the evolution of magnetic energy in astrophysical simulations by performing a standard test problem for MHD codes, Orszag-Tang. We estimate the numerical dissipation in the simulations using state-of-the-art numerical simulation code in astrophysics, \pluto. The estimated numerical resistivity in 2D simulations corresponds to the Lundquist number $\approx 10^{4}$ in the resolution of $512\times512$ grid cells. It is also shown that the plasmoid unstable reconnection layer can be resolved with sufficient resolutions. Our analysis demonstrates that in non-relativistic magnetohydrodynamics simulations, magnetic and kinetic energies undergo conversion into internal energy, resulting in plasma heating.

\end{abstract}

\begin{keywords}
Magnetohydrodynamics~-- Magnetic energy dissipation~--
resistivity~-- numerical simulations~-- PLUTO
\end{keywords}

\section{Introduction}\label{intro}
The evolution of magnetic fields has a significant impact on the dynamics and structure of astrophysical systems, ranging from stars and planets to galaxies and even the large-scale structure of the universe. Magnetic reconnection is believed to be a responsible mechanism of magnetic field evolution.
Magnetic reconnection can explain the heating and acceleration of particles or plasmoids which are observed in high energy ejections like solar flares \citep{solar1946, Jiang_2021}, magnetic substorms in the Earth magnetosphere \citep{Robert1979, Akasofu1968}, jets, and relativistic ejections from the accretion discs of compact objects \citep{BR01, Ripperda2022}.

In this study, we conduct the Orszag-Tang \citep[OT;][]{ot1}, a well-known test problem in numerical magnetohydrodynamics (MHD) simulation codes, used to examine the dissipation of magnetic energy and substructure formation in magnetized plasma. We estimate the numerical resistivity by employing a resistive MHD (Res-MHD) module in the PLUTO code \citep{m07} and we find a resolution sufficient for investigating the system's properties. Subsequently, we investigate the energy conversion in MHD simulations.

This paper is structured into the following sections:
In Section \ref{eq}, we present the MHD equations, describe the initial and boundary conditions of the OT test problem, and detail the numerical setups.
Section \ref{R} is dedicated to the discussion of simulation results, including the estimation of numerical resistivity and analysis of energy conversion. The final section provides the conclusions.
\section{Numerical approach and problem conditions}\label{eq}
\subsection{Magnetohydrodynamics (MHD) method}
The MHD method is based on the equations of conservation of mass, momentum, and energy, and Maxwell's equations as follows,

\beqa
    \frac{\partial \rho }{\partial t}+\boldsymbol{\nabla} \cdot (\rho \boldsymbol{\upsilon})=0, \label{mass_nonrel}\\
        \frac{\partial  \rho \boldsymbol{\upsilon}}{\partial t}+\boldsymbol{\nabla} \cdot (\rho  \boldsymbol{\upsilon} \boldsymbol{\upsilon} +p\boldsymbol{I} +\boldsymbol{T}_{EM})=0,
    \label{energy-momentum1_nonrel}\\
    \frac{\partial {\varepsilon}}{\partial t}+\boldsymbol{\nabla} \cdot \left[ \left( \omega + \rho \frac{\upsilon^2}{2} \right) \boldsymbol{\upsilon} + c\boldsymbol{E}\times\boldsymbol{B} \right]=0,
    \label{energy-momentum2_nonrel}
\eeqa

where $\rho$ and $\boldsymbol{\upsilon}$ are density and velocity, respectively. $\boldsymbol{I}$ represents the identity tensor, $T_{EM}$ is the electromagnetic stress tensor, $\epsilon$ is the total energy density and $\omega = p+U_{int}$ is enthalpy density, where $U_{int}$ is the internal energy. Magnetic and electric fields are denoted with $\boldsymbol{B}$ and $\boldsymbol{E}$ respectively.
\beqa
    \frac{1}{c} \frac{\partial \boldsymbol{B}}{\partial t}+\boldsymbol{\nabla}\times \boldsymbol{E}=0, \label{EM1}\\
     \frac{1}{c}\frac{\partial \boldsymbol{E}}{\partial t}-\boldsymbol{\nabla}\times\boldsymbol{B}=-\boldsymbol{J}/c, \label{EM2}
\eeqa
in the above equations, $\boldsymbol{J}$ is the current density that comes from Ohm’s law,
\beqa
     \boldsymbol{J} = \frac{c^2}{\eta}(\boldsymbol{E}+\frac{\boldsymbol{v}}{c}\times \boldsymbol{B}) = c \boldsymbol{\nabla} \times \boldsymbol{B} \ ,
    \label{current_nonrel}
\eeqa
where, in the cgs system of units we use, $\eta$ represents the physical resistivity.
\subsection{Initial and boundary conditions}
We solve the OT test problem using the MHD approximation method. This problem is a standard test problem in MHD codes to examine the power of the code to capture the MHD shocks and shock-shock interactions.

We set up the simulations of OT test problem in a 2D box $0<x,y<2\pi$ with periodic boundary conditions \citep{ot1} with the polytropic equation of stat with adiabatic index $\gamma=4/3$.

The initial velocity and magnetic fields of OT test problem are,
\beqa \label{OT01}
    &v = v_{0}(-\sin y, \sin x, 0),\\
    &B = B_{0}(-\sin y, \sin 2x, 0),
\eeqa
we define $v_{0} = 0.99 c/ \sqrt{2}$ and  $B_{0} = c \sqrt{4\pi \rho_0}$ where $\rho_0$ is the scaling factor of density in code unit. The simulation results are presented in the code unit where the factor $1/ \sqrt{4\pi}$ and speed of light $c$ are absorbed in the magnetic field and velocity, respectively. The initial density and pressure are constant ($\rho = 1$ and $P = 10$ in the code unit).
\subsection{Numerical setup}
We perform the OT test problem in Minkowski coordinate in the Newtonian code \pluto~ v.~4.4.
The equations are solved using the \texttt{HLLC} Riemann solver and the Monotonized Central difference limiter \texttt{MC}. The equations are evolved in time using the second-order Runge-Kutta method (\texttt{RK2}). To ensure $\boldsymbol{\nabla}\cdot \boldsymbol{B} = 0$, we use the flux-constrained transport \texttt{CT} method.
\section{Simulation results and discussion}\label{R}
The simulations are performed in the resolutions in the range of $64\times 64$ to $4096\times 4096$ grid cells, with the step of multiplying by 2 in each direction. All simulations run to the final time $t=10t_c$ ($t_c=L/c$ is the light-crossing time with $L=1$ as the typical length of the system).

We study the energy evolution of the system by exploring the averaged energy densities $\overline{Q}$ defined by
\beqa
    \overline{Q}=\frac{\iint_S Q \,dx\,dy}{\iint_S  \,dx\,dy}
\eeqa
\subsection{Numerical resistivity} \label{Res}
To investigate the system evolution, we determine the required resolution at which the numerical error has the least effect on the physics of the system.

In this section, we assess the numerical resistivity in simulations at different resolutions using the resistive MHD (Res-MHD) module in \pluto.
The sufficient resolution is the resolution in which the numerical resistivity is less than the physical resistivity and the substructures (plasmoids) are captured. Plasmoids are regions of higher density and lower magnetization relative to their surroundings that may exist in the magnetic reconnection layers. Plasmoids evolve along the reconnection layers by growth, bulk acceleration, and mergers. Theoretical studies show that the plasmoid unstable reconnection layers exist when the Lundquist number ($S=L/\eta$) is larger than $10^4$ \citep{BR01, Loureiro_2007}. We search for the resolution that the numerical resistivity in the code unit is less than $10^{-4}$.

\begin{figure}[h!]
    \centering
    \includegraphics[width=0.45\columnwidth]{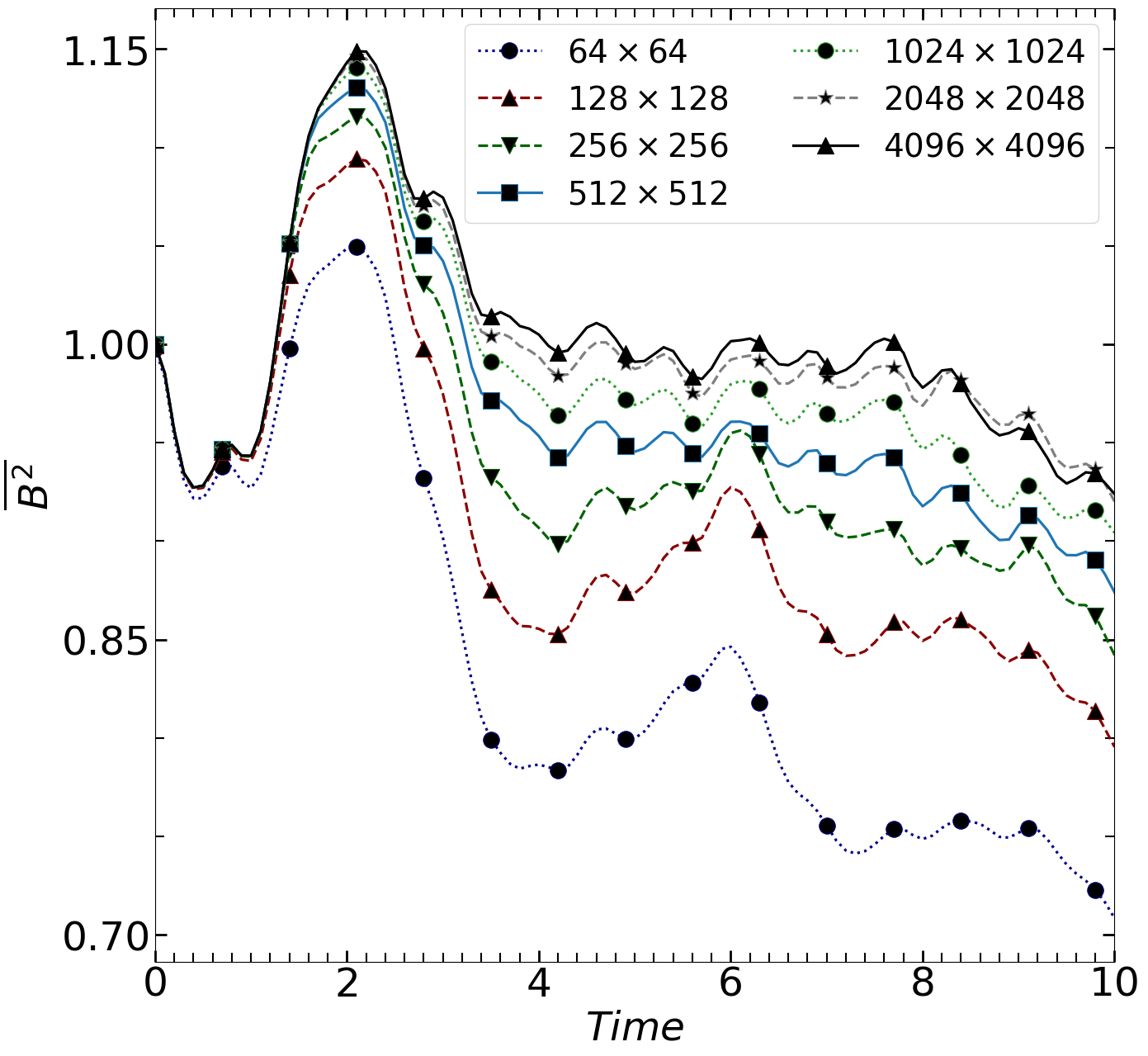}
    \includegraphics[width=0.45\columnwidth]{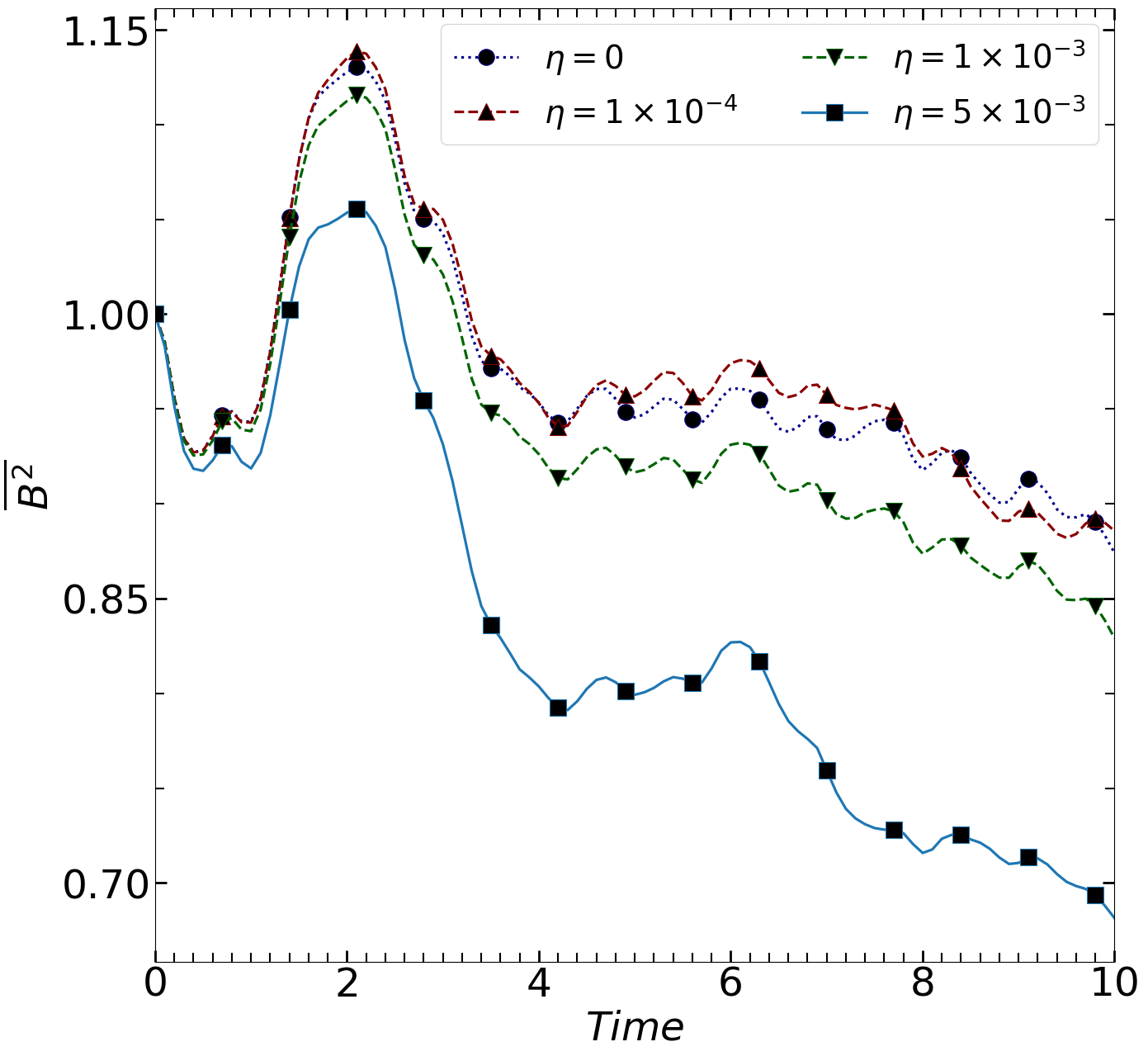}
        \caption{Time evolution of $\overline{B^2}$ in MHD simulations in different resolutions (\emph{Left}) and in Res-MHD simulations in the resolution of $512\times 512$ grid cells and different physical resistivities (\emph{Right}) using PLUTO code.}
    \label{MHD}
\end{figure}

In the left panel of Fig. \ref{MHD}, we present the time evolution of $\overline{B^2}$ in the non-resistive, ideal MHD simulations at various resolutions. The value of $\overline{B^2}$ shows an increase as the resolution is enhanced (numerical resistivity decreases). This indicates that as the resolution increases, the curves exhibit greater convergence.

To determine the numerical resistivity at each resolution, we compare the time evolution of $\overline{B^2}$ in ideal MHD simulations ($\eta=0$) and Res-MHD simulations at different physical resistivity values (e.g. $\eta=10^{-4}$, $10^{-3}$, and $5\times 10^{-3}$). When the Res-MHD curve converges to the MHD curve, it indicates that the numerical resistivity is less than/equal to the physical resistivity.

For example, in the right panel of Fig. \ref{MHD}, we present the results of simulations at a resolution of $512\times 512$ grid cells. It is evident that the curve corresponding to $\eta=10^{-4}$ converges to the MHD curve ($\eta=0$). This convergence implies that the numerical dissipation at the resolution of $512\times 512$ grid cells is $\geq 10^{-4}$. Therefore, the resolutions $\geq 512\times 512$ grid cells can be suitable for studying the evolution and properties of the system.

\begin{figure*}
\includegraphics[width=0.45\columnwidth]{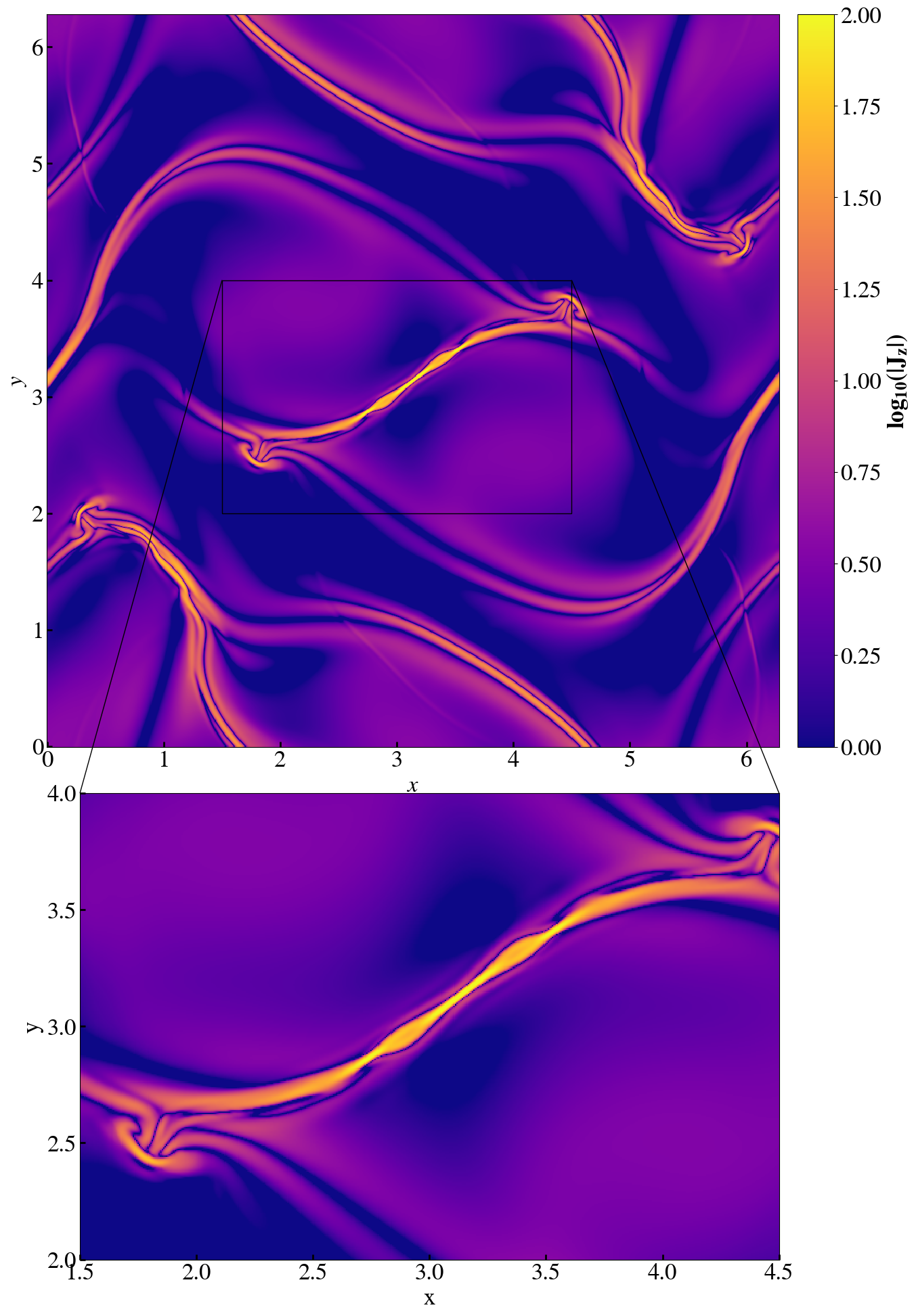}
\includegraphics[width=0.45\columnwidth]{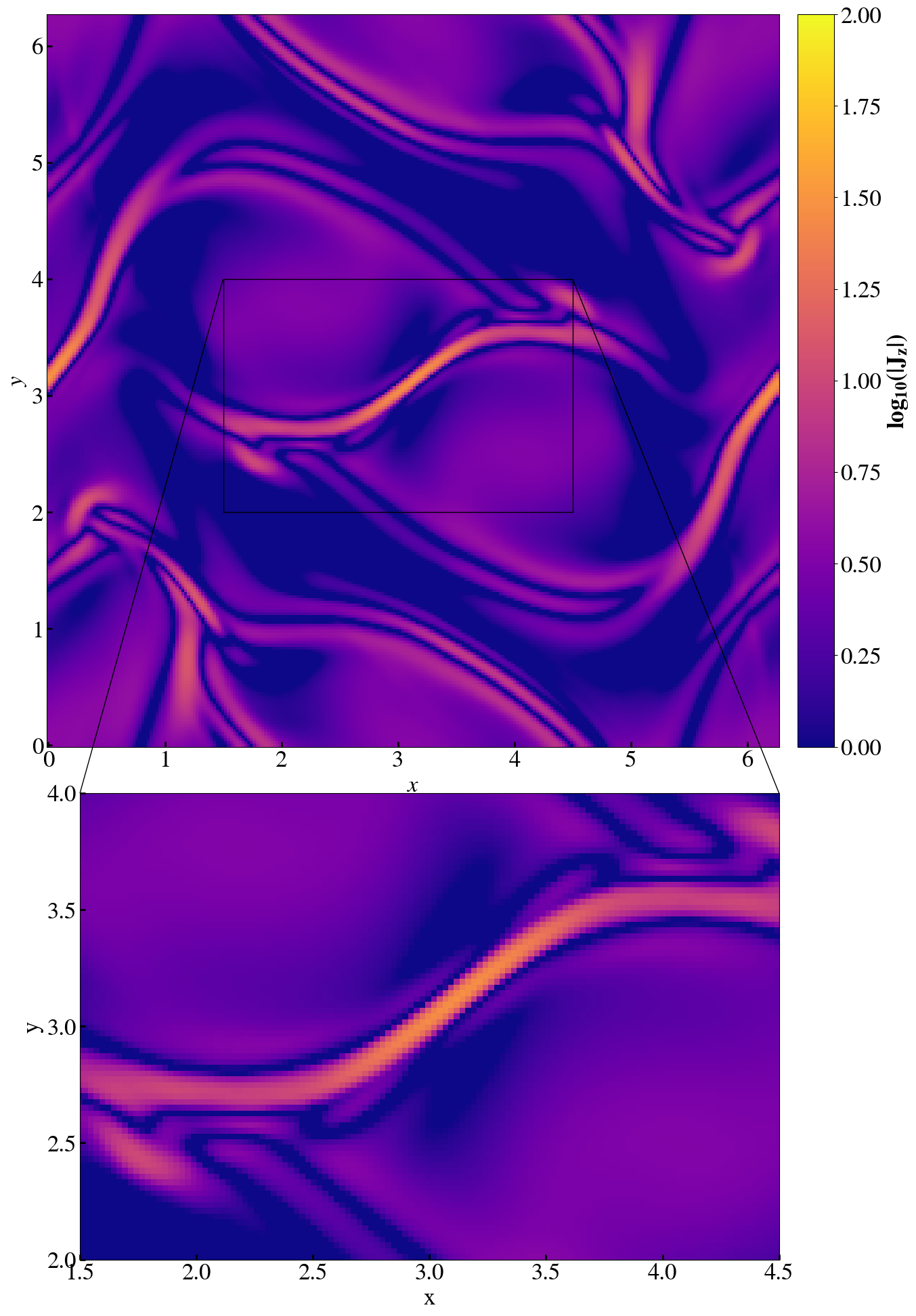}
\caption{\textbf{Current density $J_z$} at $t = 2.5 t_c$ of MHD simulation in the resolution of $1024\times 1024$ (\emph{left}) and  $256\times 256$ grid cells (\emph{right}). The bottom frames show the zoomed-out of the reconnection layer in the middle of the simulation box.}
\label{DMHD}
\end{figure*}

In Fig.\ref{DMHD} are shown the \textbf{current densities} in ideal MHD simulations in two resolutions at $t=2.5 t_c$. The zoomed frames at the bottom show the reconnection layer in the middle of the simulation boxes. The left panel shows the results with the resolution of $1024\times 1024$ grid cells, where the numerical resistivity is less than $10^{-4}$ so the current layer is thin enough to become plasmoid unstable. The right panel shows the results in lower resolution $256\times 256$ grid cells that give higher numerical resistivity ($> 10^{-4}$),  the layer is thick and plasmoids do not emerge. We studied the magnetic reconnection and plasmoid formation in different models of resistive MHD, relativistic MHD, and 3D simulations in \cite{kayanikhoo2023}. In addition,  \citet{Puzzoni2021} studied, the particle acceleration in the reconnection layer using \pluto~concerning the impact of Riemann solvers and reconstruction methods, grid resolutions, and numerical resistivity on particle acceleration in magnetic reconnection simulations. Their findings demonstrate that particle acceleration undergoes variations based on the choice of numerical solvers and Lundquist numbers ($S\geq 10^{3}$).

%

\subsection{Energy conversion in MHD simulations}\label{en}
In this section, we study the magnetic energy dissipation in MHD and Res-MHD simulations. The conserved total energy density contains magnetic energy density $E_B = B^2/2$, kinetic energy density $E_k = \rho \upsilon ^2/2$, internal energy $U_{int} = p/(\gamma-1)$, and electric energy density $E_E = E^2/2$.
\begin{figure}
    \centering
    \includegraphics[width=0.9\columnwidth]{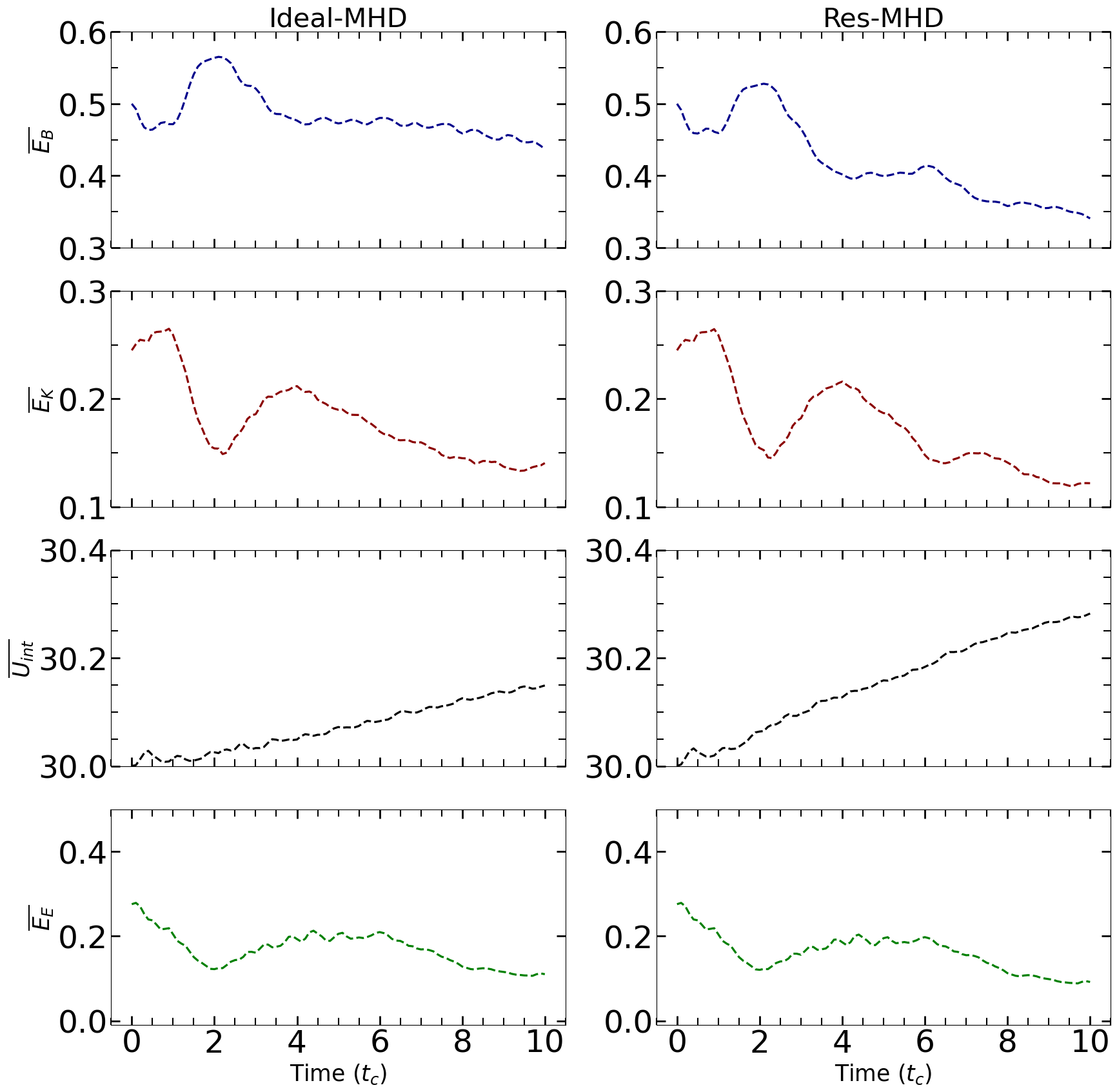}
    \caption{The time evolution of the averaged density of energy components in the simulations MHD and Res-MHD ($\eta=5\times10^{-3}$) at the resolution of $512\times 512$ grid cells.}
    \label{energy}
\end{figure}
We present the energy components that are computed at the resolution of $512\times 512$ grid cells in Fig. \ref{energy}. The left column displays ideal MHD simulations ($\eta=0$), while the right column represents Res-MHD simulations ($\eta=5\times10^{-3}$).

In the top row of panels, we show the time evolution of magnetic energy. The magnetic energy increases to $\approx1.15$ times its initial value at  $t\simeq 2t_c$, coinciding with the presence of a current layer. Shortly after, at $2.5t_c$, plasmoids emerge within the current layer in the ideal MHD simulation (shown in Fig. \ref{DMHD}). Subsequently, magnetic energy gradually dissipates as the simulation progresses. In the Res-MHD simulation, we note that the magnetic energy dissipates at a faster rate compared to the ideal MHD simulation.

In the second row of Fig.\ref{energy}, we show the average density of kinetic energy. The evolution of kinetic and magnetic energies illustrates that kinetic energy leads to the amplification of the magnetic field at approximately $2t_c$. Magnetic energy converts to kinetic energy through the magnetic reconnection to the time step $t\simeq 4t_c$.

Both kinetic and magnetic energies dissipate into internal energy by the end of the simulation time, heating the plasma, as shown in the third-row panels. It is evident that resistivity results in increased dissipation, causing internal energy to rise by approximately 0.5\% more than in the ideal MHD simulation.
The last row of panels represents electric energy, which evolves similarly to magnetic and kinetic energy, as expected.


\section{Conclusions}\label{conclus}
 In this paper, we investigated the magnetic energy evolution in the magnetized plasma by performing the Orszag-Tang (OT) test problem, in astrophysical magnetohydrodynamic (MHD) code \pluto~v.~4.4. OT is a standard test problem to test the power of MHD codes to capture shocks and resolve the substructures.

 Using the resistive MHD module in \pluto, we estimate the numerical dissipation in each resolution. The numerical resistivity in the resolution of $512\times 512$ grid cells is estimated to be
 $\geq 10^{-4}$ which represents the lowest limit for the existence of the plasmoid-unstable current layer. In the ideal MHD simulation in the resolution of $1024\times 1024$ grid cells, the plasmoid-unstable current layer is observed while in the lower resolution, $256\times 256$ grid cells the plasmoids do not appear. Furthermore, our results confirm that in the Res-MHD simulation with a physical resistivity of $10^{-4}$, plasmoids do not emerge within the layer.

We studied the energy evolution in MHD and Res-MHD simulations. Our findings show that kinetic energy drives the amplification of the magnetic field, resulting in magnetic energy reaching approximately $1.15$ times its initial value. In MHD simulation, shortly after this increase, the plasmoids are observed within the current layer. Subsequently, both magnetic and kinetic energies gradually dissipate over the course of the simulation, with the dissipated energy converting into internal energy and heating the plasma. The results indicate that in resistive MHD simulation with resistivity $\eta = 5\times10^{-3}$, the dissipation is $0.5\%$ higher than in MHD simulation.

In \cite{kayanikhoo2023}, we further examined 3D simulations, compared relativistic and non-relativistic MHD simulations through energy conversion and magnetic reconnection rates, and the impact of resolution on reconnection rates in the models. Additionally, we compared two MHD codes widely used in astrophysics \pluto~and \koral.

\ack
This project was funded by the Polish NCN grant No.2019/33/B/ST9/01564. M\v{C} acknowledges the Czech Science Foundation (GA\v{C}R) grant No.~21-06825X. MW was supported by the European Research Council advanced grant “M2FINDERS - Mapping Magnetic Fields with INterferometry Down to Event hoRizon Scales” (Grant No. 101018682). High-resolution computations in this work were performed on the Prometheus and Ares machines, part of the PLGrid infrastructure.
\bibliography{OTKragtbib}

\end{document}